\documentclass[fleqn,usenatbib]{mnras}

\usepackage{newtxtext,newtxmath}
\usepackage[T1]{fontenc}
\usepackage{ae,aecompl}

\usepackage{graphicx,subfigure}	% Including figure files
\usepackage{float}
\usepackage{caption}
\usepackage{amsmath}	% Advanced maths commands
\title[Short title, max. 45 characters]{Revisiting profile instability of PSR~J1022+1001}

\author[P. V. Padmanabh et al.]{
P. V. Padmanabh,$^{1}$\thanks{E-mail: prajwalvp@mpifr-bonn.mpg.de}
E. D. Barr,$^{1}$
D. J. Champion,$^{1}$
R. Karuppusamy,$^{1}$
M. Kramer,$^{1,2}$
\newauthor
A. Jessner,$^{1}$
P. Lazarus$^{1}$
\\
$^{1}$Max Planck Institute for Radio Astronomy, Auf dem H\"ugel 69, Bonn 53121, Germany\\
$^{2}$Jodrell Bank Center for Astrophysics, University of Manchester, Alan Turing Building, Oxford Road, Manchester M13 9PL, United Kingdom\\
}

\date{Accepted XXX. Received YYY; in 
13
 original form ZZZ}

\pubyear{2020}

\begin{document}
\label{firstpage}
\pagerange{\pageref{firstpage}--\pageref{lastpage}}
\maketitle
\graphicspath{{plots/}}

\begin{abstract}

Millisecond pulsars in timing arrays can act as probes for gravitational wave detection and improving the solar system ephemerides among several other applications. However, the stability of the integrated pulse profiles can limit the precision of the ephemeris parameters and in turn the applications derived from it. It is thus crucial for the pulsars in the array to have stable integrated pulse profiles. Here we present evidence for long-term profile instability in PSR~J1022+1001 which is currently included in the European and Parkes pulsar timing arrays. We apply a new evaluation method to an expanded data set ranging from the Effelsberg Pulsar Observing System back-end used in the 1990s to that of data from the current PSRIX backend at the Effelsberg Radio Telescope. We show that this intrinsic variability in the pulse shape persists over time scales of years. We investigate if
systematic instrumental effects like polarisation calibration or 
signal propagation effects in the interstellar medium causes the
observed profile instability. We find that the total variation cannot be fully accounted for by instrumental and propagation effects. This suggests additional intrinsic effects as the origin for the variation. We finally discuss several factors that could lead to the observed behaviour and comment on the consequent implications.

\end{abstract}


%%%%%%%%%%%%%%%%%%%%%%%%%%%%%%%%%%%%%%%%%%%%%%%%%%

%%%%%%%%%%%%%%%%% BODY OF PAPER %%%%%%%%%%%%%%%%%%

\section{Introduction}

Millisecond pulsars (MSPs) show properties that serve as an ideal platform for testing general relativity in the strong field regime  \citep[e.g][]{2006Sci...314...97K}, probe the equation of state for neutron stars  \citep[e.g.][]{Lattimer_2001,2013Sci...340..448A} and may allow to detect gravitational waves in the nanohertz regime  \citep[e.g.][]{1990ApJ...361..300F}. Key to these investigations is the technique of pulsar timing where the arrival time of the pulsar signal is monitored by comparison against a reference template of the average pulse profile. Here, typically, tens of thousands of pulses are averaged to form an integrated profile. A template is generated from such integrated profiles across many observations. In order to avoid self imaging effects \citep{Hotan_2005}, the grand average profile is often converted into a template by modelling the pulse as a superposition of multiple Gaussians or von Mises functions. This noise free template is then used as a reference and cross correlated with profiles from each epoch to calculate the times of arrival (TOAs). An accurate timing model of the spin and and binary parameters of the pulsar should yield timing residuals which follow a Gaussian distribution with zero mean. However, the residuals most commonly show systematic variations due to several sources of noise that could be deterministic or stochastic in nature \citep[e.g][]{Cordes_2013}.

The method described above assumes that the integrated profile is invariant with time for a given observing frequency. However, several pulsars are known to show variable integrated pulse shapes. Some variations are stochastic. For example, when a few pulses (tens to hundred) are combined to form a profile, the phase jitter introduced causes noise in the timing residuals. This is mainly caused due to individual pulses fluctuating in shape, flux as well as phase. Longer integration times tend to average out these changes and form much more stable profiles, though some variability may still persist at some level due to other effects. Changes may arise from the integrated profile switching between two or or more profiles termed as "moding". These transitions typically switch back within a few thousand pulses \citep[e.g. ][]{Backer_1970,Rankin_2013}. Some pulsars have been observed to show
variations in pulse profiles on longer time scales of weeks, months or years \citep[e.g. ][]{Lyne_2010} which is longer than typical moding timescales.

Several studies have tried to  characterise jitter \citep[]{Oslowski_2011,2012MNRAS.420..361L,Shannon_2014}, intermittency \citep[e.g. ][]{Mottez_2013} as well as moding \citep{Cordes_pulsar_2013}, but the underlying mechanism that induces the transition between stable configurations of the pulse shape is not understood. Other variations are secular, being especially prominent in some binary pulsars over timescales of years. These arise from relativistic spin precession \citep[e.g.][]{Kramer_98,Desvignes_2019,VVK_2019} due to misalignment of their spin with the orbital angular momentum. This effect can be modelled and helps constraining the pulsar beam's structure.

MSPs are generally known to have a stable average pulse profile
\citep[e.g. ][]{Shao_2013}. Their short spin period allows many pulses to be averaged on time scales of minutes. Any change in the average profile of a pulsar can cause subtle unexpected variations in the timing residuals. This is due to the varying fiducial point while calculating TOAs. Devising techniques that can cope with profile changes are important to resolve this issue. On the other hand, studying these variations carefully can also give a deeper insight into factors that could impact the pulse shape. These can be broadly categorised into instrumental and non-instrumental effects. To understand these factors  in more detail, we have a conducted a detailed study of PSR~J1022+1001, for which past analyses have reported contradictory results regarding the profile stability of this pulsar \citep[]{Kramer_1999,Ramachandran_2003,2004MNRAS.355..941H,2013ApJS..204...13V,Liu_2015,2016AcASn..57..517S}.

In Section 2, we review previous profile stability studies of PSR~J1022+1001. In Section 3 we explain the procedure for data reduction of the different data sets as well as the analysis tools used. In Section 4 we revisit the procedures and methods for characterising pulse profile stability. Section 5 discusses the results from applying  new techniques to various data sets. We also cover the impact of polarisation miscalibration as well as
interstellar scintillation effects. Section 7 discusses the scientific explanations for the results obtained. We state our conclusions in section 8.

\section{Previous profile stability studies of J1022+1001}

PSR~J1022+1001 is a binary pulsar with a spin period of 16.45~ms in a 7.8-day orbit \citep{1996ApJ...469..819C} with a CO white dwarf companion \citep{1996ASPC..105..497L}. The average pulse profile exhibits a two peaked structure at 1.4 GHz (see Figure \ref{sample_profile}).
The pulsar is currently a part of the European Pulsar Timing Array (EPTA) and the Parkes Pulsar Timing Array (PPTA) among other pulsars in the quest to detect nanohertz gravitational waves. Several studies across many years using different telescopes have shown conflicting results regarding the pulse profile stability of this pulsar.

\cite{1996ApJ...469..819C} reported unusual pulse profile variations on the time scale of a few minutes within the first few months of timing this pulsar with the Arecibo telescope. \cite{Kramer_1999} observed this pulsar with the Effelsberg radio telescope and found profile changes across integration times of approximately 40 min.  Given that the shape changes persist even after integrating for hundred thousand pulses, \cite{Kramer_1999} suggested that moding cannot explain the observed instability.

Turbulence in the interstellar medium causes constructive or destructive interference leading to enhancing or weakening of the propagating pulsar signal in time as well as observing frequency. This is termed as scintillation. This propagation effect 
coupled with  profile evolution across the observing bandwidth could lead to changes in the integrated pulse profile \citep{Ramachandran_2003}.
However, \cite{Kramer_1999}  observed profile changes within a narrow frequency band (40~MHz) that could not be explained by scintillation effects. The calculated TOAs showed that the profile variation impacted the timing stability of this pulsar. The root-mean square (r.m.s) residuals post fitting were approximately 15-20 $\mu$s.

Altitude-Azimuth (Alt-Az) telescopes have to account for the rotation of the feed with respect to the sky as given by the parallactic angle during observations. Any miscalibration in polarisation could lead to leakage in power between the polarisation channels. This  would in turn affect the shape of the profile if the pulsar is highly polarised as PSR J1022+1001. To confirm that the seen changes were not telescope and mount dependent, \cite{Ramachandran_2003} observed this pulsar with the equatorial mount Westerbork Synthesis Radio Telescope (WSRT), as an equatorially mounted telescope is not affected by the change in parallactic angle. \cite{Ramachandran_2003} observed shape variation in time thus confirming the results obtained by  \cite{Kramer_1999}. \cite{Ramachandran_2003} also hypothesised that the shape changes are possibly due to magnetospheric return currents and profile variation.

Subsequent studies were carried out by \cite{2004MNRAS.355..941H} using the Parkes observatory. They noticed that the integrated profile was remarkably stable and showed no significant signs of instability with time. In contrast to the results obtained by \cite{Kramer_1999}, they obtained a much smaller r.m.s in timing residuals of 2.27 $\mu$s. They also observed that calibrated profiles showed higher degree of instability in comparison to uncalibrated profiles. While this may appear surprising, it may indicate that instrumental errors could possibly contribute to profile variations.

\cite{Purver_2010} conducted another analysis of profile variations in PSR~J1022+1001. He used data from Effelsberg, Parkes as well as the WSRT to study possible profile changes on timescales of 10 min. He also analysed profiles with a bandwidth of 8~MHz to ensure minimal impact from profile evolution with frequency. The reduced chi-squared values of his shape change parameters indicated significant changes in the pulse profile. He argued that these changes are not biased by the metrics themselves. He also argued that any instrumental errors that could cause the shape change would be well in excess of the two \emph{per cent} change in total intensity as reported by \cite{2004MNRAS.355..941H}

\cite{2013ApJS..204...13V} suggested that instrumental polarisation artefacts could lead to systematic errors in the timing residuals. He observed that cross-coupling variations over 7 years of data with the Parkes telescope introduced systematic errors close to 1~$\mu$s in TOA estimation. He developed a new calibration technique and used PSR~J1022+1001 as a test bed and demonstrated that the weighted r.m.s residuals could be reduced to 880~ns, a factor of at least two lower than previous analyses. The integration time used was roughly one hour. Through simulations, \cite{2013ApJS..204...13V} suggested that \cite{Kramer_1999} observed the variations due to changes in the parallactic angle resulted in mixing of the total intensity with the linearly polarised power of the profile. He also commented that the same variation seen by \cite{Ramachandran_2003} can be attributed to inaccurate calibration of the polarimetric response of the system.

\cite{2012MNRAS.420..361L} analysed five MSPs including PSR~J1022+1001 using parameterisation of the profile shape and phase jitter to analyse the profile stability. This analysis suggested that the profile does not vary significantly with time. However, they also pointed out that the shape parameter used in their analysis describes the entire on-pulse region, and may not be sensitive to profile variations occurring exclusively at the peaks.

More recently, \cite{Liu_2015} carried out single pulse as well as integrated profile studies using simultaneous observations from Effelsberg and Westerbork. They investigated effects of potential polarisation miscalibration of the Effelsberg data by using the template matching technique \citep{WvS_2006}. After modelling the differential gain and cross coupling of the feeds, \cite{Liu_2015} showed that applying different models to calibrate the data accounts for only a marginal fraction of the observed profile variability. They showed that subpulses from the leading and trailing components of the pulse are correlated. They also showed that subpulse variation is not correlated with the integrated profile variability observed.

\cite{2016AcASn..57..517S}  analysed Parkes data taken from the PDFB4 backend. The relatively large bandwidth of this system (256~MHz) allowed for studying pulse profile changes across frequency. The data were polarisation calibrated using the method described in \cite{Wvs_2004}. They observed that the pulse profile changed across the bandwidth. They also noticed that the  flux density for each frequency subband was different. They suggested that the profile variations are a consequence of profile evolution with frequency coupled with interstellar scintillation.

While all the studies have shown methodologically robust ways of characterising the pulse profile stability, the conflicting results provide little clarity on the origin or even the existence of variability in this pulsar. The profile variability observed could result from ISM effects in the form of scintillation or miscalibration of the instrument or due to phenomena intrinsic to the pulsar and its magnetosphere.

\section{Observations and Data Reduction}

To understand and seek to resolve the differences seen by the aforementioned analyses, we collected data that span close to two decades. We chose the data sets used by  \cite{Kramer_1999} and \cite{2004MNRAS.355..941H} given that their results disagree with one another. \cite{Kramer_1999} used data from the Effelsberg radio telescope recorded with the Effelsberg Pulsar Observing System (EPOS) \citep{1996lara.conf..185J}. 
\cite{2004MNRAS.355..941H} used data from Parkes recorded with the CPSR2 backend \citep{2005ASPC..328..377H} which generates coherently dedispersed profiles. We extended the time baseline of the data set by using data from the Effelsberg telescope from the latest PSRIX backend \citep{Lazarus_2016}. Table \ref{data set_table} provides exact details of the span of each data set used.

\begin{table}
    \centering
    \caption{Different data sets used as well as their year spans for this analysis. (N) indicates the number of epochs used for analysis}

    \begin{tabular}{||p{1.1cm}|p{1.1cm}|p{1.25cm}|p{0.7cm}|p{1.25cm}||}
\hline
\textbf{Telescope} & \textbf{Backend} & \textbf{Span} & \textbf{N} & \textbf{Dedispersion} \\ 
\hline\hline
Effelsberg & EPOS & 1995-2005 & 50 & Incoherent \\ \hline
Parkes & CPSR2 & 2004-2010 & 19 & Coherent \\ \hline
Effelsberg & PSRIX & 2011-2015 & 25 & Coherent \\ \hline

\end{tabular}%
\label{data set_table}
\end{table}

The data presented here correspond to the L-band radio wavelengths (1.3 - 1.4 GHz) with the receivers tuned to slightly different centre frequencies. The data were homogenised through clipping to a common overlapping band centred at 1390 MHz with a 40 MHz bandwidth.
The EPOS backend was connected to an online incoherent dedisperser along with an adding polarimeter and did not provide frequency resolution \citep{refId0}. Thus, we averaged across the entire clipped bandwidth for all epochs of the CPSR2 and PSRIX data to maintain consistency across data sets. We also optimised for any smearing caused by an inconsistent ephemeris by refolding the archives with the best known parameters after fitting for a linear drift in the pulse across the observed time.

Reduction of the archive files was done using the  \texttt{PSRCHIVE} \footnote{http://psrchive.sourceforge.net/} package \citep{Hotan_2004}. The profiles were manually cleaned to mitigate Radio Frequency Interference (RFI) using \texttt{pazi}. A few epochs from CPSR2 showed inaccurate linear polarisation and circular polarisation both before and after calibration. These epochs turned out to have no signal from one of the polarisation channels for every alternate sub integration of an epoch. Data showing these artefacts were removed from the analysis to avoid inconsistencies.

The \texttt{PSRCHIVE} Python module was used to extract data from archive files before applying methods and techniques to characterise the pulse profile shape. The details of these methods are discussed in the following section. Archive files with signal-to-noise ratio (S/N) below 100 were excluded from the analysis to avoid uncertainties limited by poor S/N. The data sets from the CPSR2 and PSRIX backends were polarization calibrated using 
\texttt{pac}\footnote{http://psrchive.sourceforge.net/manuals/pac/}. The Jones matrix solutions for the Parkes multibeam were derived using observations from PSR~J0437-4715 based on the measurement equation template matching (METM) technique as described in \citep{2013ApJS..204...13V}. We calibrated the PSRIX data using the single axis model \citep{Britton_2000}.

To use a uniform analysis environment, we modified the current \texttt{PSRCHIVE} plugin\footnote{https://sourceforge.net/p/psrchive/code/ci/master/tree/Base/Formats/EPOS/} for the legacy EPOS data. Subsequently we used the \texttt{PSRCHIVE} Python plugin for the rest of the analysis. We used a Python based pipeline ( \texttt{EPOS\_calibrator} \footnote{https://github.com/prajwalvp/EPOS\_calibrator}) in order to calibrate the EPOS data. The pipeline implements calibration based on equations for an adding polarimeter as specified in \cite{refId0}.

\section{Methods}

\subsection{Previous methods}

\cite{Kramer_1999} used a simple but straight forward metric to describe the pulse profile stability of PSR~J1022+1001. They used the amplitude ratio computed from the two peaks in the integrated profile and took a ratio between them with one of the peaks as a reference. If the profile is stable, one would expect the amplitude ratio to be stable across time. They observed that a profile integrated up to 10-16 min could show different amplitude ratios with a 30 \emph{per cent} scatter.

\cite{2004MNRAS.355..941H} used a different metric to characterise the profile stability. Every integrated profile spanning 5 min was first normalised based on the flux density in the on-pulse region. A high-S/N normalised profile was chosen as a standard reference template. The template was then subtracted from the normalised profile for every epoch. All the subtracted residuals were stacked and the standard deviation was calculated per phase bin across epochs. If the profile were stable with time, the residuals should behave like white noise and the standard deviation across phase bins should ideally remain constant across the  profile. This result reported in \cite{2004MNRAS.355..941H}, indicates no sign of instability. 

\subsection{Peak optimiser}

In order to first investigate the data with a method similar to the first method in the previous subsection, we apply a robust approach that uses specific features of the profile. We focus on the peaks of the two components of the pulse profile. The integrated profile is generated for every epoch such that the on-pulse is roughly centered and not impacted by the pulse profile wrapping in phase. To ensure consistency across all data sets, we generate integrated pulse profiles with 1024 bins across rotational phase.

\begin{figure}
    \centering
    \includegraphics[scale=0.35]{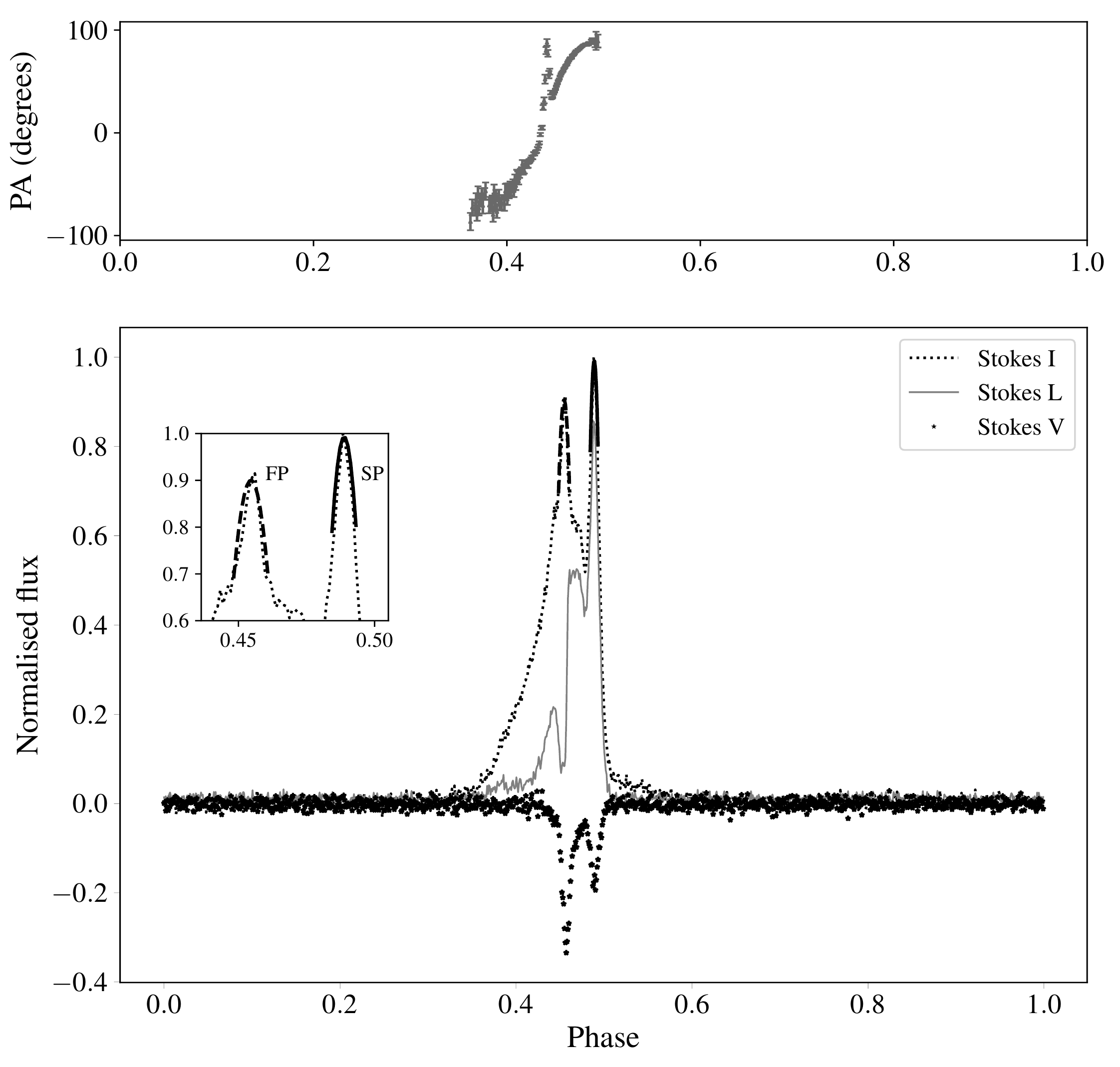}
    \caption{Bottom panel shows a sample Stokes profile for PSR~J1022+1001 at 1.4 GHz observed with CPSR2. The plot inset shows an example second order polynomial fit on the first peak (FP) and second peak (SP). The top panel shows the position angle (PA) as a function of rotational phase. }
    \label{sample_profile}
\end{figure}

Since PSR~J1022+1001 has two well defined peaks in its profile at 1.4 GHz (see Figure  \ref{sample_profile}), we choose the first peak (hereafter FP) and the second peak (SP) as features to analyse the stability of the pulse profile.  We next fit for both phase and amplitude of FP and SP.

Firstly, the phase bin with the maximum amplitude is used as a starting reference for each of the peaks. A window centered on the reference point is chosen to represent the peak. We then solve for the turning point of a Least-Squares second order polynomial fit. The fit has the advantage of yielding a single maximum and does not over-interpret the shape of the profile. It also requires fewer points compared to a higher degree polynomial fit. A different window size is chosen for FP and SP since they differ in shape and sharpness of the peak. The size of the selected window for each peak is chosen by using the reduced  ${\chi}^2$ . For this, we chose different window sizes and applied the above mentioned method. We then calculate the reduced ${\chi}^2$ value for the obtained fit with respect to the original profile in the window of interest. We eventually choose the window size that yielded the reduced ${\chi}^2$ value closest to 1. This yielded a window size of 20 bins for FP and 10 bins for SP.

To make a robust determination of the uncertainties in the fit, we take a Monte Carlo-based approach. We calculate the off-pulse r.m.s and generate multiple realisations of the selected window per peak with data in the window drawn from a Gaussian distribution with mean as the value of the point within the window and the off-pulse r.m.s as the standard deviation for that point. We then apply the parabolic fit for each realisation and solve for the phase bin value at which the turn over of the fit occurs. The final reported value comes from the median of the distribution of the peaks obtained from the different iterations. The error bars reported come from choosing the values at the 5th and 95th percentile of the distribution.

To make this procedure more robust, we use the new phase value as a starting point reference with the same previously selected window size per peak. The Monte Carlo approach is applied once again to solve for the maximum. This bootstrapping technique helps improve the  characterisation of the peak. We also apply a similar procedure to characterise the amplitude. Once the windows are finalized and the fit is applied, we solve for the amplitude keeping the phase bin value fixed. This fixed phase bin value comes from solving for the maximum from the second order fit as mentioned previously. Once the phases and amplitudes for each of the peaks have been solved for, we use two different metrics to characterise the profile stability. Firstly, we take a difference of the solved phase bin values for FP and SP. Secondly, we take an amplitude ratio of the solved amplitude values between both the peaks similar to \cite{Kramer_1999}.
We will refer to the phase difference and amplitude ratio calculation together as the Peak Optimiser (hereafter PO).

\subsection{Modified Hotan method}

In addition to these two metrics referring to phase and amplitude, we also apply a slight modification to the method used by \cite{2004MNRAS.355..941H}. The original method uses a high S/N reference profile template and stacks the difference amplitudes of profiles with respect to the template from every epoch. The standard deviation per phase bin of the stacked profiles is also calculated to make this more robust. We modify this method and calculate a median of all average profiles across all chosen epochs to use as a reference template for subtraction. This makes the analysis less susceptible to statistical outliers. We align each profile in phase before subtraction. This is done by rotating the profiles in the Fourier domain and additionally optimising for fractional phase bin variation if any. To avoid residual errors from normalisation, we also model for an offset in the on-pulse region of the standard deviation per phase (STDP) plot and apply a correction using a cost function to minimise the r.m.s of the STDP. We will hereafter refer to this method as MHM.

%\subsection{KS test}
\vspace{3mm}

In order to compare data between backends, we have applied the Kolmogorov-Smirnov test \citep{ks_test}. The advantage of this test is that it is non parametric. We select data pair wise between the three backends in this analysis and test the null hypothesis that the two distributions in consideration come from the same parent distribution.

\section{Results}

Using the new and modified stability metrics as discussed in the previous section, we analyse data from three backends (see Figure \ref{mkm_full}). Within each data set, we distinguish between polarisation calibrated and uncalibrated data. Polarisation calibration is essential in eliminating systematic effects. For instance, incorrect calibration can affect the two detected polarisation channels differently and eventually result in a Stokes I variation. \cite{2013ApJS..204...13V} suggested that miscalibration had a major effect on precision pulsar timing for several pulsars including PSR~J1022+1001 and so developed a technique that modelled possible artefacts in order to account for this. We thus expanded the analysis to study the effect of improper polarisation calibration on the pulse profile stability across all the data sets. We discuss the results obtained on each data set below.   

\subsection{EPOS}

The EPOS data span approximately 10 years from 1995 to 2005 and the observation lengths for all epochs range between 20 and 30 min. The results from applying PO and MHM are shown below.

\begin{figure*}
    \centering
    \includegraphics[scale=0.4,keepaspectratio]{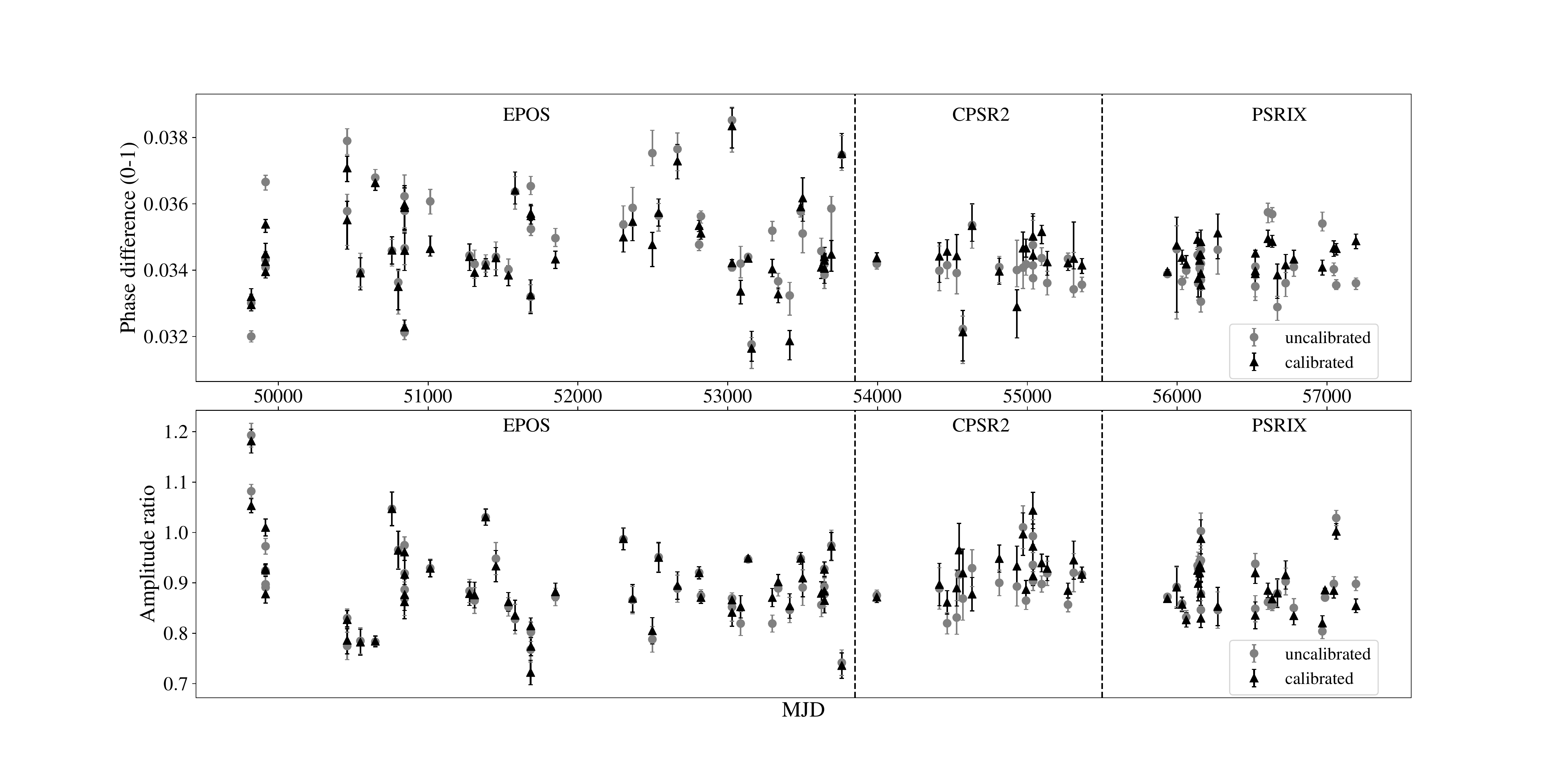}
    \caption{(a) PO results for all three data sets across 20 years. Top panel shows the difference in peak phases plotted against time. The bottom panel shows amplitude ratios between peaks plotted against time.}
    \label{mkm_full}
\end{figure*}
%]

Figure \ref{mkm_full} shows the results of applying PO to the EPOS data set. It is seen that there is significant scatter in both phases and amplitude ratios signifying profile instability. While calibration reduces the scatter marginally (1-2 \emph{per cent}), there is still unaccounted variability observed for the profile across time.  \cite{Kramer_1999} had observed a 30 \emph{per cent} scatter in the amplitude ratio across an observation and we observe a slightly higher ( 38 \emph{per cent} ) scatter across a long range of epochs.

The results from applying MHM can be seen in figures \ref{hotan_method_epos_uncal} and \ref{hotan_method_epos_cal}). The standard deviation around the on-pulse region is high for uncalibrated (21-sigma significant) and calibrated data sets (16-sigma significant). The difference amplitude plots also show clear variations in the on-pulse region. Hence, the figures clearly indicate that the profile is unstable on timescales of 20-30 min which is commensurate with the results from applying the PO.

\begin{figure}
    \centering
    \includegraphics[scale=0.35]{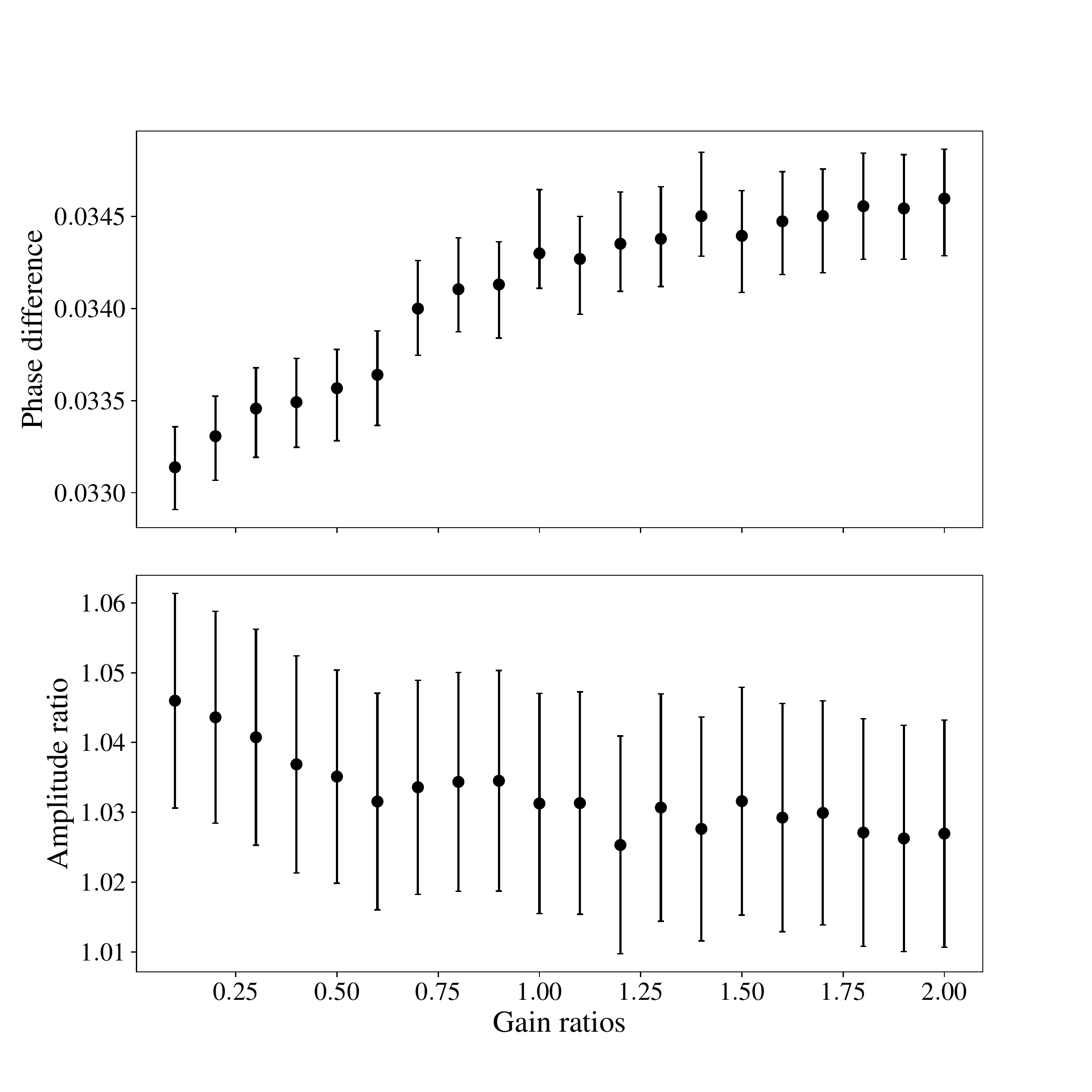}
    \caption{Top panel: Plotted is the difference in phase of the two peaks for a sample EPOS profile against the ratio in the gains for the two polarisations. Bottom panel shows the amplitude ratio as a function of the gain ratio.}
    \label{epos_gain}
\end{figure}

To assess the effect of polarisation calibration, we varied the ratio of the gains of each polarisation and applied PO on each gain ratio (see Figure \ref{epos_gain}). The gain levels in each polarisation are typically within 10 dB  of each other. Figure \ref{epos_gain} shows that the range of phase differences seen by changing the gain ratios from 0.1 to 2 is close to 0.002 (in phase). This is at least a factor of 2 smaller than the range of phase differences observed in Figure \ref{mkm_full}. Similarly, the amplitude ratio changes by about 0.05 which is close to a factor of 10 smaller than the change observed in Figure \ref{mkm_full}. This indicates that although we change the gain ratio by more than an order of magnitude, it cannot fully account for the observed variability in the integrated pulse profile. This provides strong evidence that instrumental gain effects cannot fully explain the profile instability observed for the EPOS data.

\subsection{CPSR2}

The Parkes CPSR2 data span approximately 7 years from 2004 to 2010 and were obtained from the CSIRO Data Access Portal\footnote{https://data.csiro.au/dap}. The observation lengths chosen were between 20 and 30 min. 

\begin{figure}
    \centering
     \includegraphics[scale=0.45]{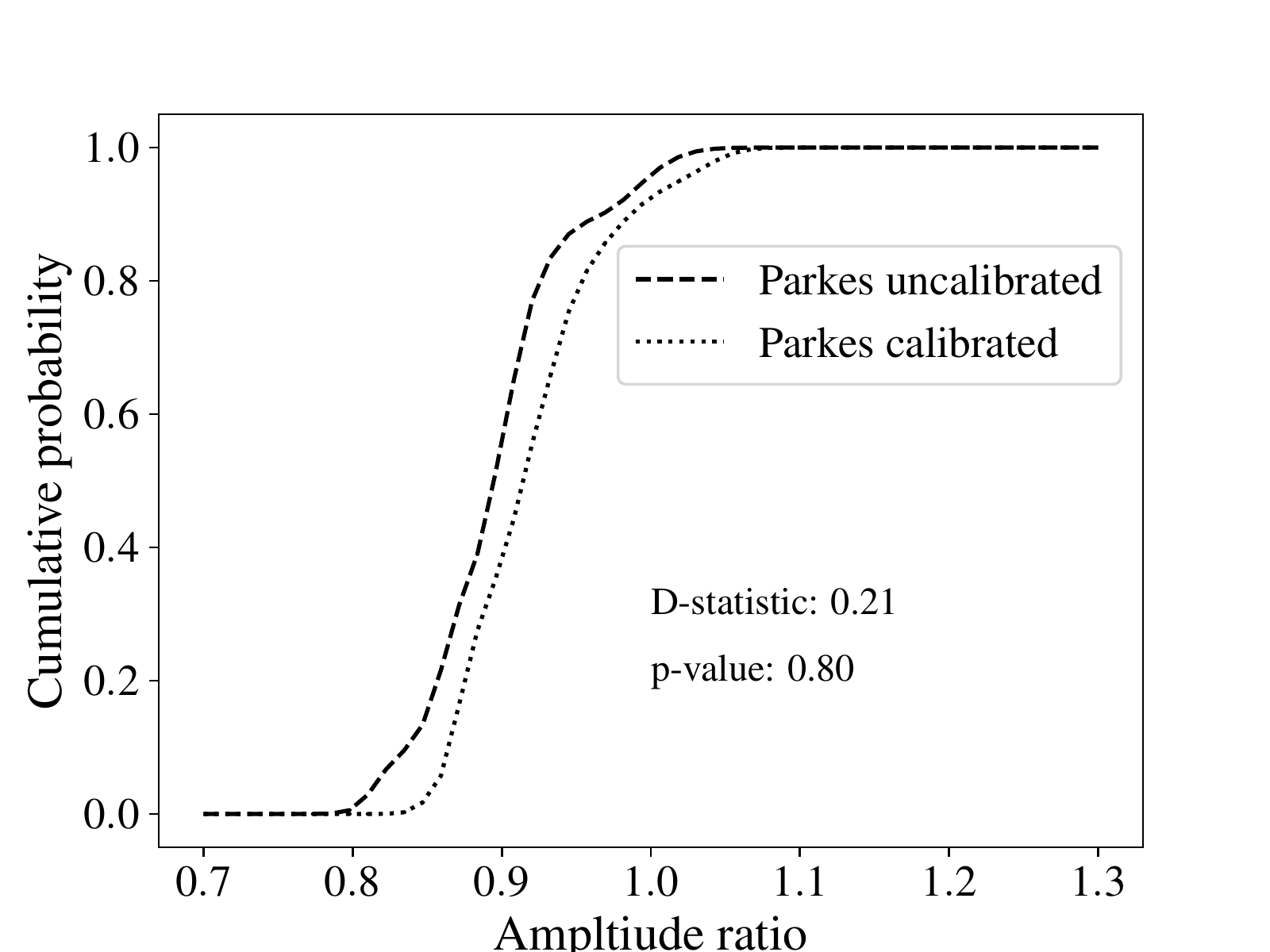}
         \caption{Comparison of cumulative distribution functions for CPSR2 data between uncalibrated and calinrated data.}
    \label{ks_test_parkes}
\end{figure}

Figure \ref{mkm_full} shows that the scatter in phase difference increases by about 0.002 units when the data are calibrated. But we observe that calibration helps to reduce the scatter on the amplitude ratio. We further investigated the impact of miscalibration by inspecting the cumulative distribution function for the calibrated and uncalibrated data. We focused on the amplitude ratio given that this parameter is less susceptible to artefacts compared to the phase difference (as seen from the polarisation analysis in the previous subsection). We observe a scatter of roughly 30 \emph{per cent} (see Figure \ref{ks_test_parkes}) for the cumulative distributions. We also observe a slight offset between the distribution functions of the uncalibrated and calibrated Parkes data sets . However, this difference is not significant to ascertain that calibration makes an impact (as indicated by the p-value).

In order to investigate if the profile is unstable at smaller timescales than 20-30 min, we first applied the original method used by \cite{2004MNRAS.355..941H} on their data with an integration length of 5 min (See Appendix \ref{appendix} for details). The results for the uncalibrated data are consistent with the results from \cite{2004MNRAS.355..941H}. We also compared the results for applying the original method on the shorter integration (5 min) as well as longer integration (20-30 min). We observe that the calibrated as well as uncalibrated data for the long integrations show significant variability and the calibration is observed to improve the profile stability across epochs.     

Apart from applying the original method, we also applied the MHM on the data set used by \nopagebreak{\cite{2004MNRAS.355..941H}}. The difference amplitudes show no clear evidence for instability but the standard deviation of all epochs across the rotation phase of the pulse show a near 4-sigma significance in the on-pulse region. 

\begin{figure*}
    \subfigure[EPOS uncalibrated]{\includegraphics[scale=0.3]{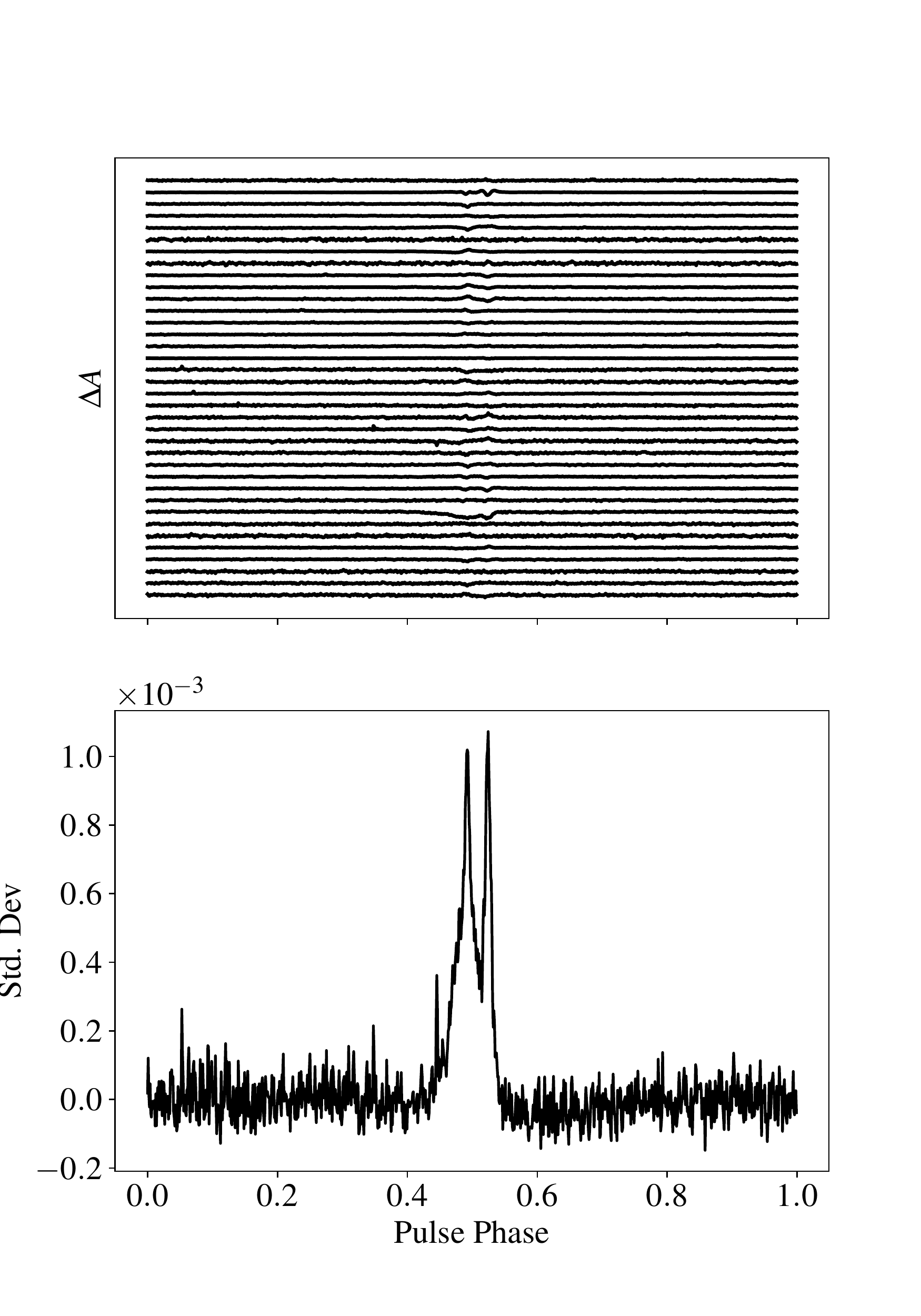}
    \label{hotan_method_epos_uncal}}
    \subfigure[CPSR2 uncalibrated]{\includegraphics[scale=0.3]{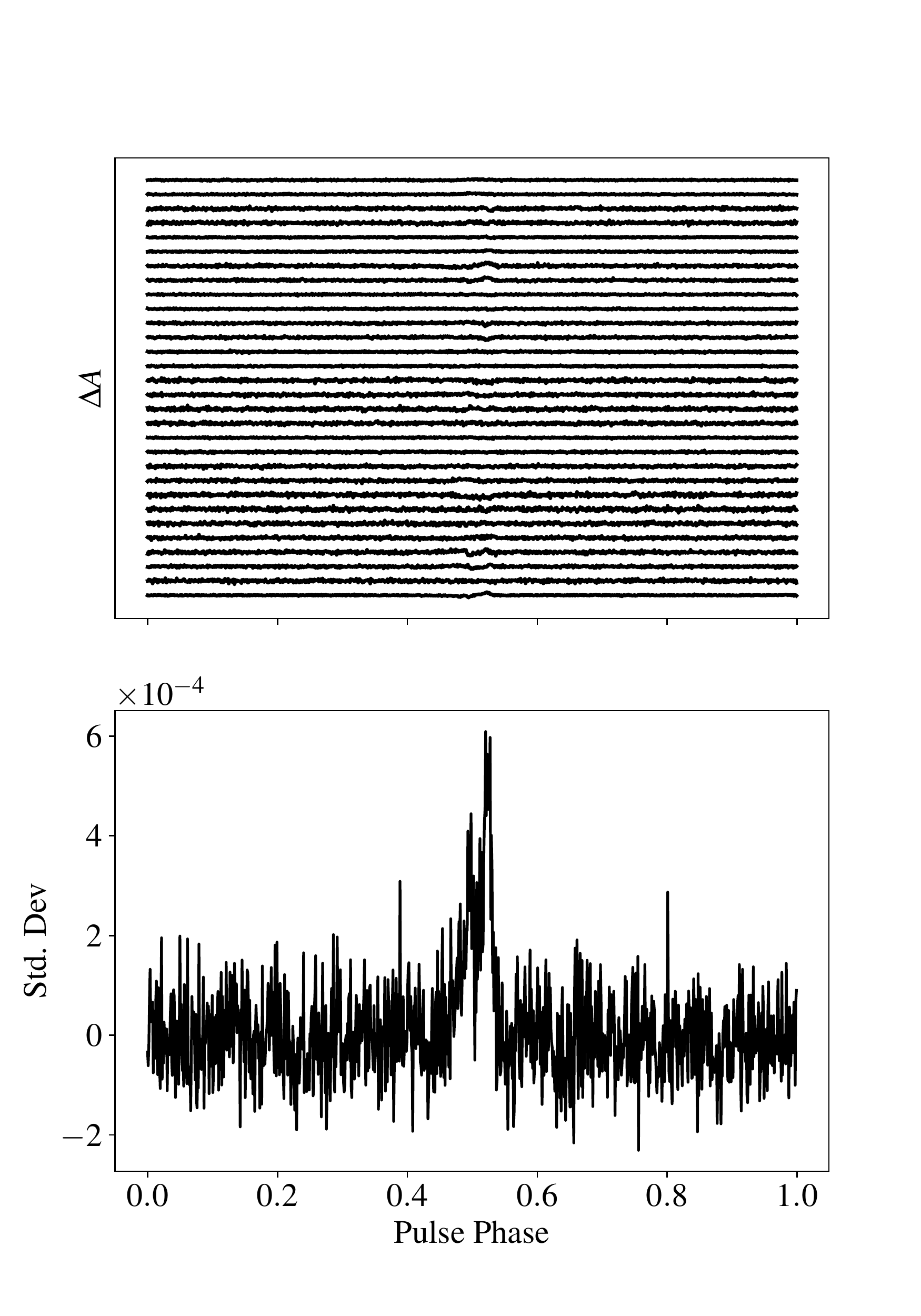}
    \label{hotan_method_parkes_uncal}}
    \subfigure[PSRIX uncalibrated]{\includegraphics[scale=0.3]{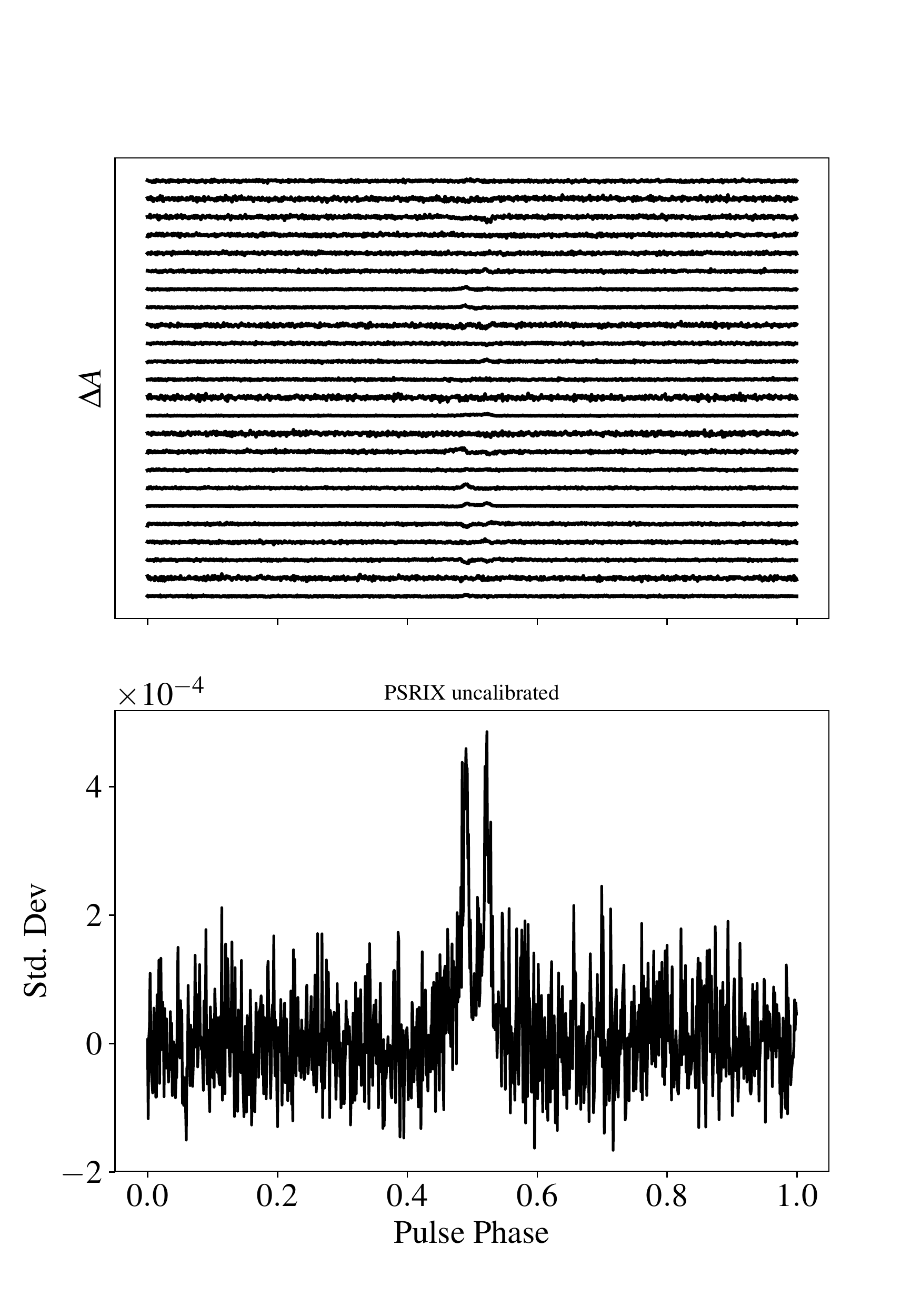}
    \label{hotan_method_psrix_uncal}}
    \subfigure[EPOS calibrated]{\includegraphics[scale=0.3]{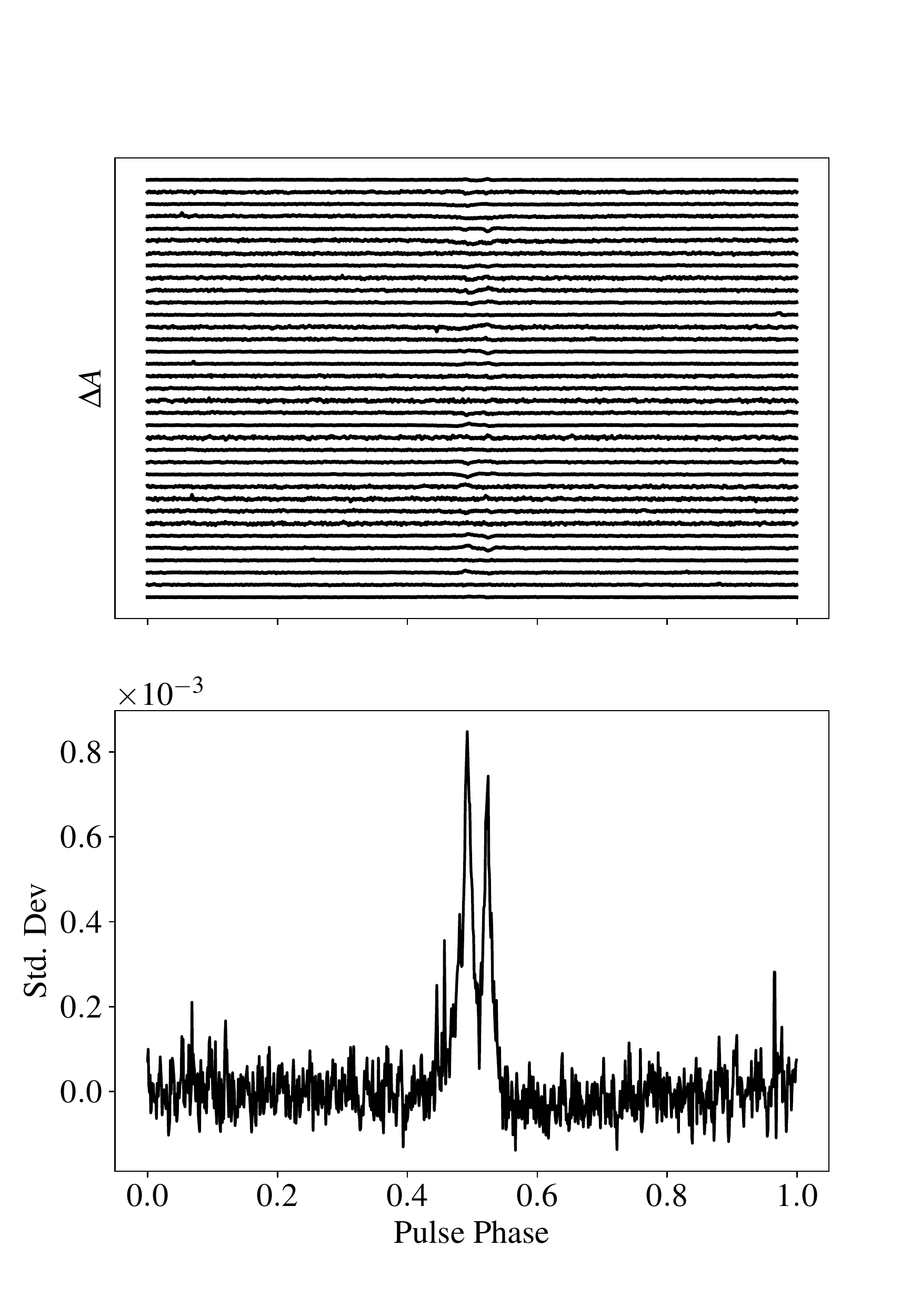}
    \label{hotan_method_epos_cal}}\quad 
    \subfigure[CPSR2 calibrated]{\includegraphics[scale=0.3]{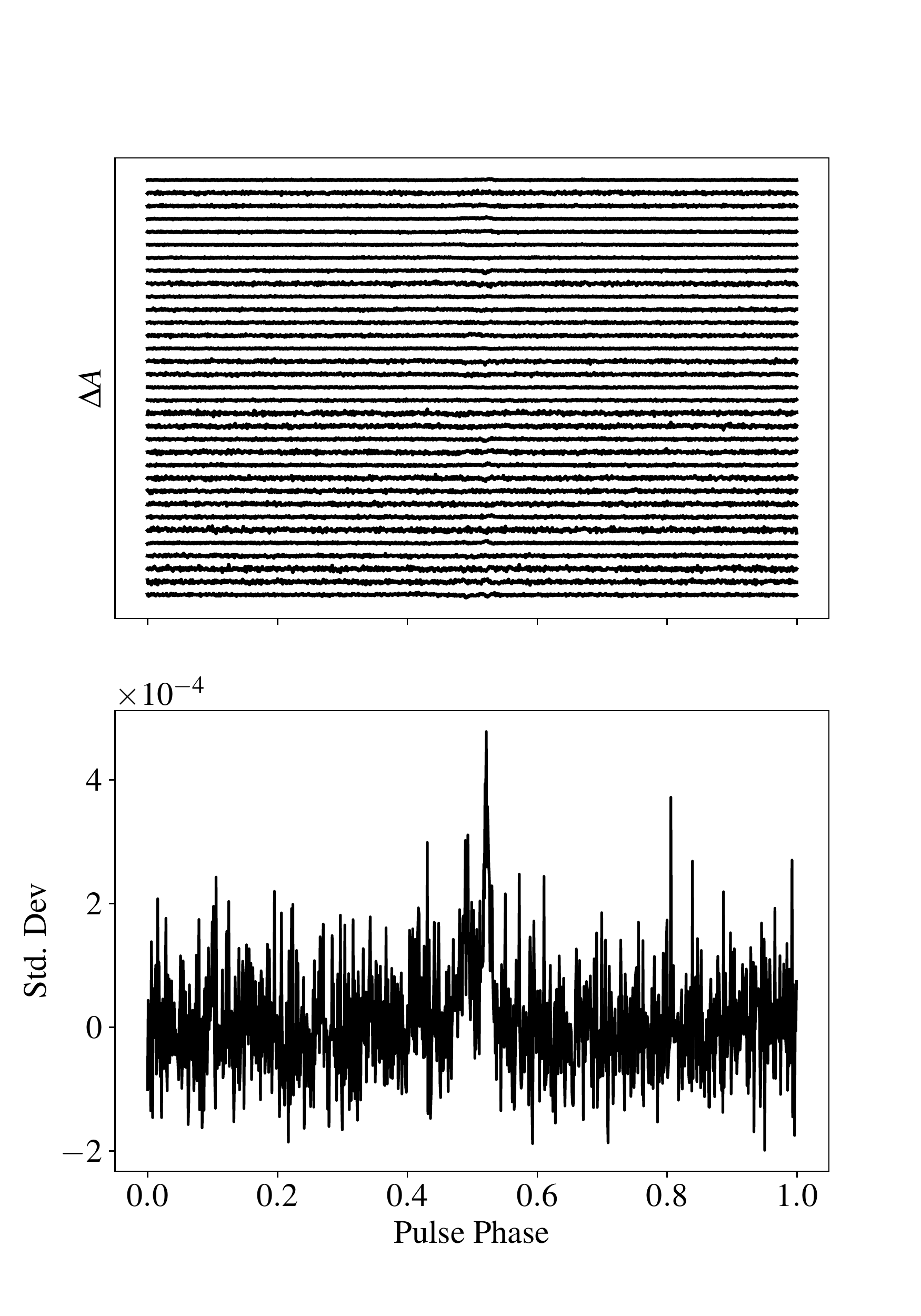}
    \label{hotan_method_parkes_cal}}\quad
  \subfigure[PSRIX calibrated]{\includegraphics[scale=0.3]{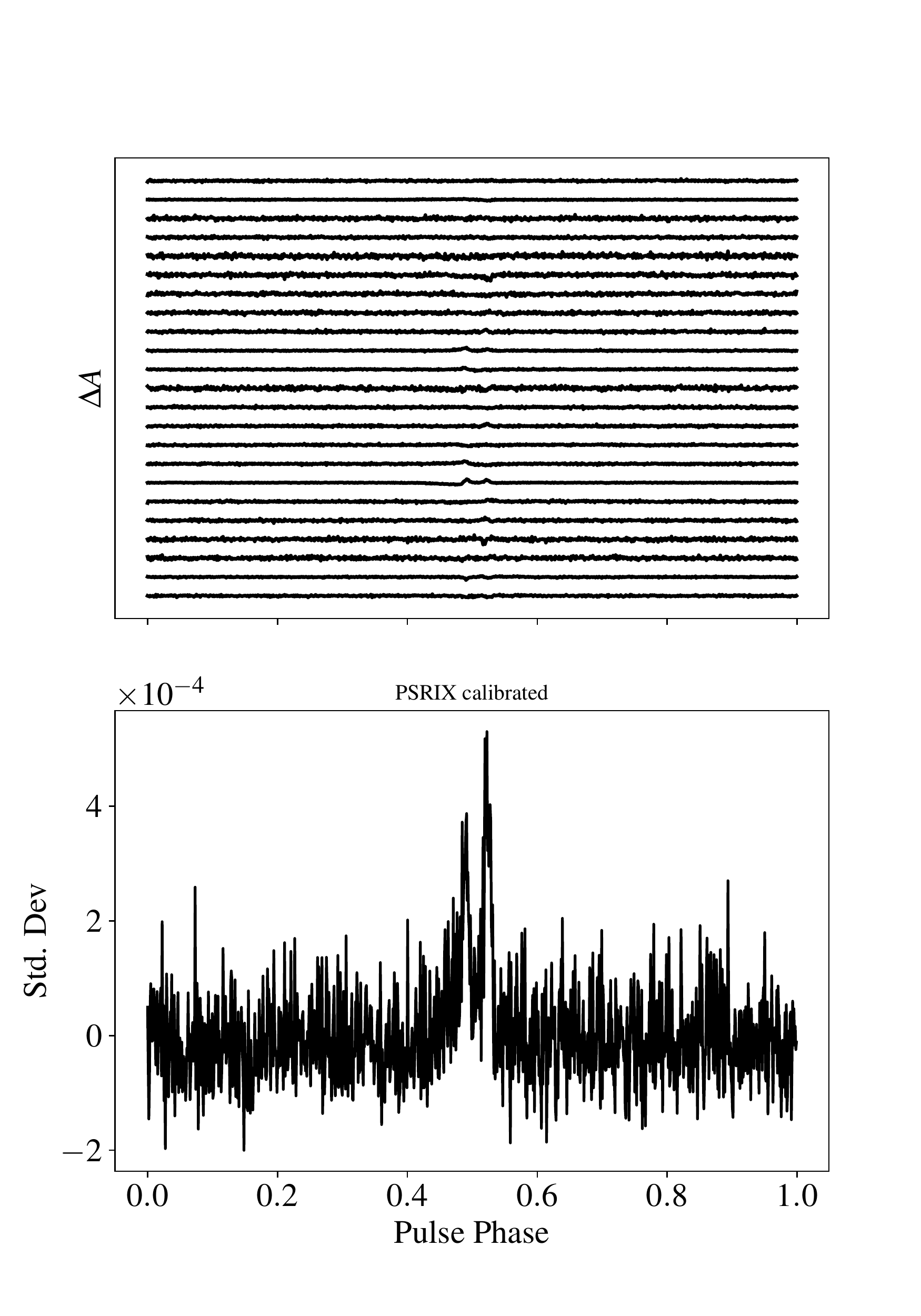}
     \label{hotan_method_psrix_cal}}
    \caption{MHM results for all calibrated and uncalibrated data. The top panel in each individual plot is denoted by $\Delta A$ which refers to the difference amplitudes stacked for flux normalised profiles. The bottom panel is the standard deviation across all epochs as a function of the rotational phase.} 
\end{figure*}

The PO analysis showed some ambiguity about the stability indicators for the 20-30 min  integration data. Thus, using a different metric would help in throwing some more light on the issue. Figures \ref{hotan_method_parkes_uncal} and \ref{hotan_method_parkes_cal} show results from applying MHM to the long integration data. The maximum standard deviation around the on-pulse region is 8-sigma above the off-pulse deviation for the uncalibrated data. This value is close to 6-sigma for the calibrated data set. These values are much more significant compared to the value obtained for integration length of 5 min. This indicates that different integration timescales can play a crucial role in these studies. The dependence on integration timescales could plausibly explain the difference between this analysis and that of \cite{2004MNRAS.355..941H}.
\renewcommand{\thesubfigure}{\alph{subfigure}}

\begin{figure}
    \centering
    \includegraphics[scale=0.3]{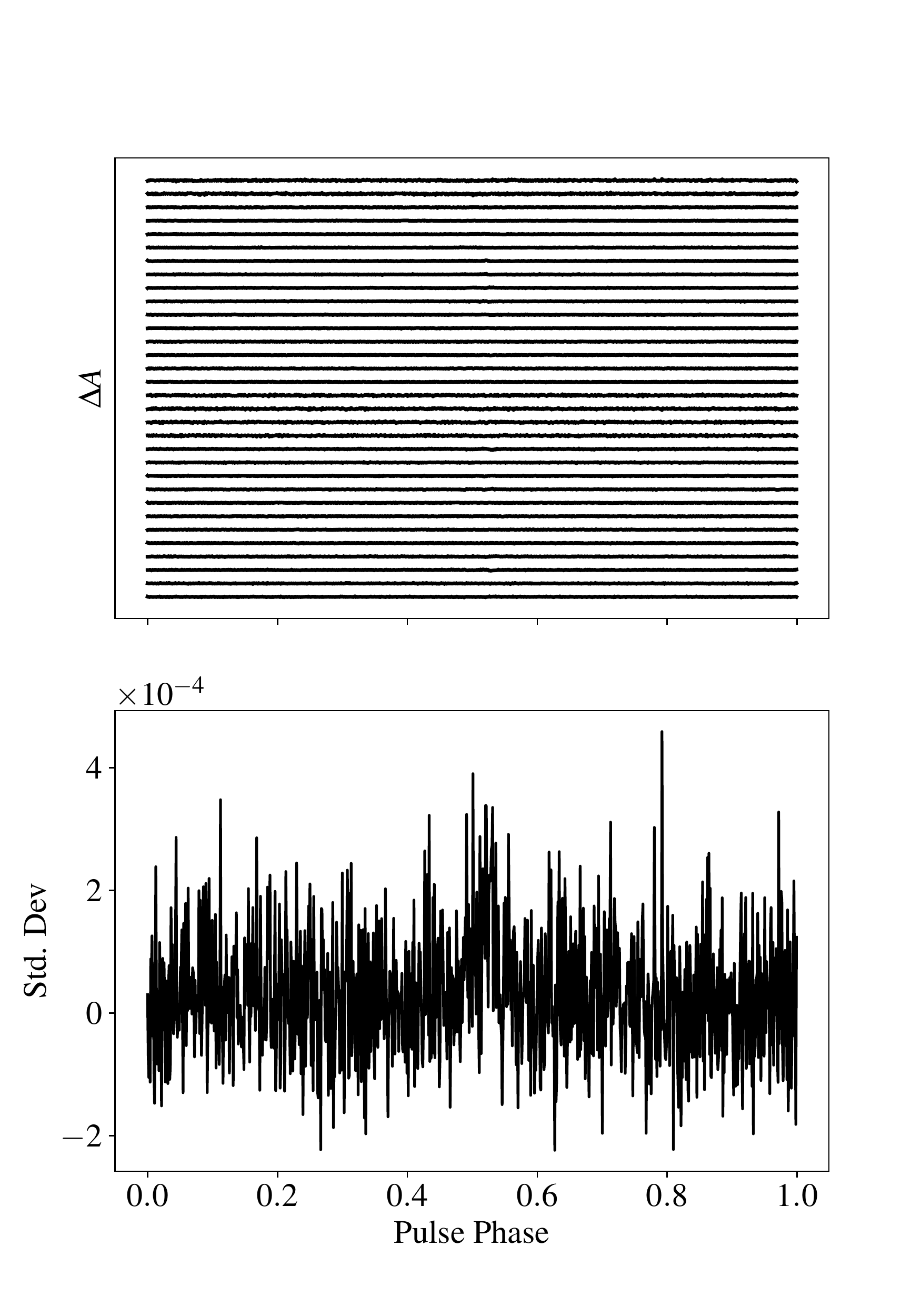}
    \caption{MHM plot for the data set used by . The integration time for each epoch is five minutes. The top panel is the difference amplitudes stacked and the bottom panel is the standard deviation per phase bin across all epochs.}
    \label{hotan_method_parkes_uncal_5min}
\end{figure}

\subsection{PSRIX}

The data taken with the PSRIX backend of Effelsberg span approximately 5 years from 2011 to 2016 with observation lengths ranging between 20 and 30 min per epoch, similar to previous data sets.

Figure \ref{mkm_full} shows results from applying the PO on the PSRIX data set. We observe that the scatter in phase differences as well as the amplitude ratio  lowers on calibration. Still, both plots support the hypothesis that the profile is unstable on these timescales as seen from the previous two data sets.

The results from applying MHM to the PSRIX data are consistent with results from applying PO (See figures \ref{hotan_method_psrix_uncal} and \ref{hotan_method_psrix_cal}). 
We observe a significant standard deviation difference  between the maximum deviation and the off pulse deviation for both calibrated (7-sigma significant) and uncalibrated (7-sigma significant) data sets. The difference amplitude plots complement the observations of the standard deviation.

\subsection{Comparing telescope data set distributions through KS test}

In order to investigate different backends,  we observe the trends in the cumulative distribution functions obtained for the amplitude ratios. This is a robust indicator of any instrumental effects that have impacted our pulse profile analysis.

\begin{figure}
    \centering
    \includegraphics[scale=0.5]{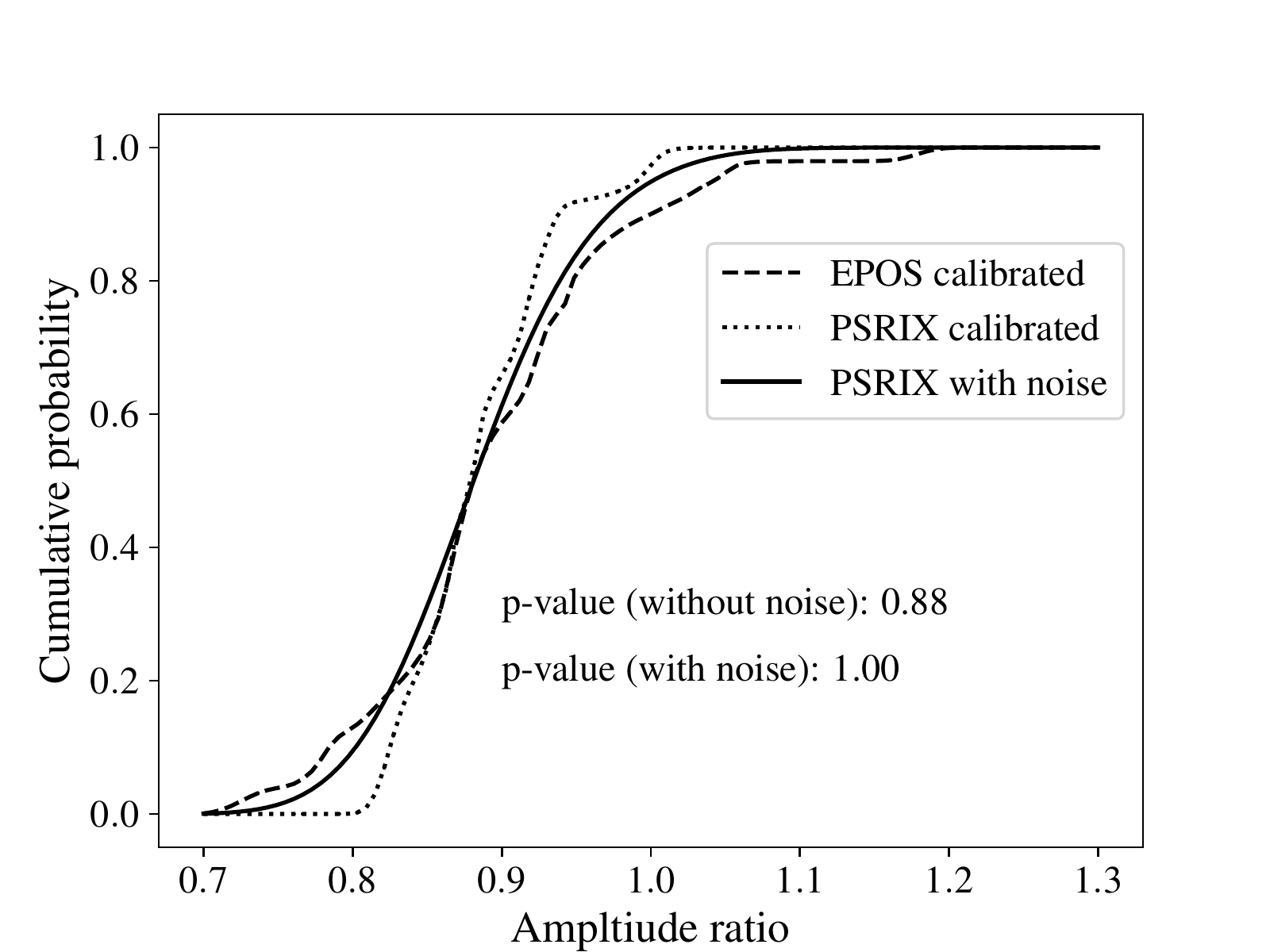}
    \caption{Comparison of cumulative distribution functions of amplitude ratios between PSRIX, EPOS data and PSRIX data with white noise.}
    \label{ks_test_epos_noise}
\end{figure}

\begin{figure}
    \centering
    \includegraphics[scale=0.5]{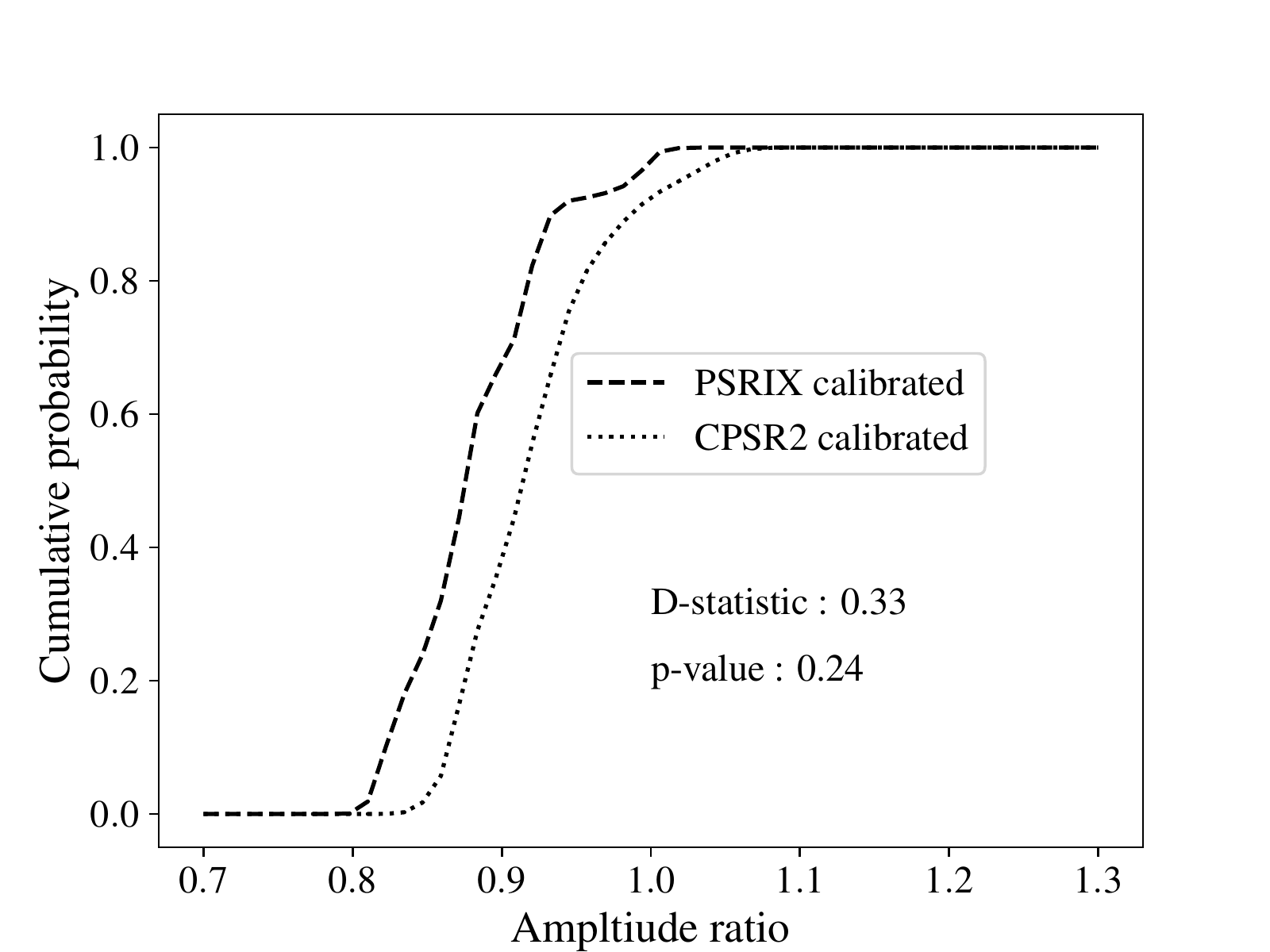}
    \caption{Comparison of cumulative distribution functions of amplitude ratios between PSRIX and CPSR2 data after calibration.}
    \label{ks_test_parkes_offset}
\end{figure}

Ideally, one would expect the distributions in pulse shape changes to be independent of the telescope backend, but as seen in figure \ref{mkm_full}, we see a narrow range of amplitude ratios for PSRIX and a much wider range for EPOS. On injecting a white noise source with nearly 4 times the r.m.s of the ampltiude ratios of the PSRIX data, we are completely able to explain the differences observed (see Figure \ref{ks_test_epos_noise}). The p-value which was 0.86 became 1.0 after injection of noise. This suggests that some part of the  EPOS backend chain significantly increased the noise floor of the incoming signal.

We also conducted a KS test between CPSR2 and PSRIX data post-calibration to check for similarities or differences. In Figure \ref{ks_test_parkes_offset}, we observe an offset between the distributions. However, the p-value indicates that we cannot reject the null hypothesis that the  PSRIX and CPSR2 distributions arise from the same parent distribution.

\section{Scintillation based analysis}

Some studies \citep{2016AcASn..57..517S,Ramachandran_2003} had suggested that profile evolution with frequency along with interstellar scintillation could lead to varying pulse profiles. \cite{Liu_2015} used simulations based on real profile evolution across frequency and showed that scintillation can introduce significant profile variations for a large bandwidth (150-200 MHz). If the profile shows different structures across different observing frequencies, scintillation can lead to variations in the integrated profile. We calculated the expected scintillation bandwidth for this pulsar using the NE2001 model \citep{cordes2001}. This is 57~MHz at 1.4~GHz which is of the order of the bandwidth used in our analyses. The scintillation timescale from the model is close to 15~minutes.  

We investigated the impact of diffractive scintillation on CPSR2 and PSRIX data. We generated median profiles for six frequency sub bands based on observations from all epochs for a particular backend. We then modelled the gain variation across the observing bandwidth as a sinusoid and shifted it in phase in steps of a quarter wavelength across the band to create multiple templates. We overlaid these templates across the bandwidth. This would weight each subband differently based on the sinusoidal template used. We finally applied the PO for the frequency scrunched profile based on each sinusoidal template. We further extended this procedure for multiple sinusoidal frequencies ranging from 0.5 to 2.5 wavelengths in steps of 0.25. We then calculated the maximum possible scatter value that can be obtained by varying the sinusoidal frequency as well as shifting the template in phase. 

We observe that the phase difference has a maximum variation of 0.0008 as compared to 0.0025 as seen in our analysis with PO for PSRIX data (see Figure \ref{mkm_full}). Thus, the variations from diffractive scintillation accounts for roughly up to 30 \emph{per cent} of the variation observed with PSRIX.  Similarly, the amplitude ratio has a maximum variation close to 25 \emph{per cent} of the total variation observed for PSRIX data. We carried out the same analysis with the CPSR2 data and the maximum variation in phase difference and amplitude ratios account up to 15 \emph{per cent} and 20 \emph{per cent}, respectively. These results indicate that scintillation cannot explain the observed instability.

\section{Discussion}

We were able to reproduce the result from \cite{Kramer_1999} for the scatter seen in the amplitude ratios across time.  Though the scatter observed in \cite{Kramer_1999} was across an observation of a few hours, our analysis showed that integrated profiles of the order of 30 min still show the same behaviour. The EPOS backend specifically shows a higher degree of instability compared to the other data sets (about 20 \emph{per cent} higher scatter). This is especially apparent from the KS test analysis where introducing a constant noise source to the PSRIX data fits the model for EPOS. At the beginning of the observation, the channel gains for EPOS were set manually on dials trying to keep the total power readings for each channel within a prescribed range typically at a medium to low elevation and higher system noise. To make sure that the manual setup is not causing any artefacts, we varied the gains of each cross polarisation up to 10 dB and reconstructed the profile for various gain ratios. The analysis showed that the profile shape change is negligible compared to the scale of changes noticed for pulse profiles across time. The dynamic range of the EPOS system was limited and interference at times saturated one of the polarisation channels and not the other. A possibility is that the first component of the pulse could saturate the receiver causing the system to be unstable and in turn leading to varying amplitudes in the second component. However, pulsar signals observed with EPOS were usually not strong enough to saturate the receiver. Additionally, if this were the case, we would not expect to see the profile shape changing for the other backends. We reproduced the results from \cite{2004MNRAS.355..941H} for 5 min integrations of the pulse profile (as shown in the Appendix). Though we notice no trend in the difference amplitudes, the standard deviation plot reveals a hint of instability that is further confirmed when applying MHM on the 5 min integrations. This slight difference in the results in comparison with \cite{2004MNRAS.355..941H} is most likely due to a slight difference in the epochs used for analysing. However, it is well evident from MHM and the original method used in \cite{2004MNRAS.355..941H} that the profile is unstable at longer timescales on the CPSR2 data. We also explored the possibility of profile evolution with frequency playing a role in shape changes. While there is clear evidence of profile evolution with frequency for this pulsar \citep{Ramachandran_2003,2016AcASn..57..517S}, this cannot directly explain the profile changes seen across time if we observe across the same range of observing frequencies. One can expect profile evolution with frequency to have an impact when a 'scintle' passes through the observing band. We thus modulated the band with a wide range of simulated scintillation frequencies and phases to determine the maximum variation that could be produced. We find that scintillation can only account for 25 \emph{per cent} of the observed variability.

The KS test analysis on the CPSR2 data set revealed that  calibration introduces a slight offset but does not play a significant role in distorting the pulse shape. In addition to the offset, we observed a scatter in the range of amplitude ratios for the CPSR2 data similar in scatter range observed by \cite{Kramer_1999} indicating that some unexplained causes of instability remain. The KS test between EPOS and PSRIX indicated the strong possibility of an increased noise floor with EPOS. Previous tests have been conducted comparing the non-dedispersing backend with the standard pulsar backend of EPOS, using a standard noise diode signal. There was a significant increase (up to a maximum factor of 4) observed, which explains the KS test result.

Previous studies have suggested several phenomena intrinsic to the pulsar which could explain the changes seen in the profile. \cite{Kramer_1999} in their analysis suggested that mode changes alone cannot explain the variability seen. We note, however,
that since this initial discussion, pulsars have been discovered to also exhibit significant profile changes on much longer timescales
(see \citep{Lyne_2010} or \citep{Karastergiou_201}). For the current analyses, we have used pulse profiles of the order close to $10^5$ pulses where one expects the pulse profile to be stabilised if ``classical mode changing'' is happening (i.e.~typically of the order of tens of minutes at most).  However, the MKM analysis (see Figure \ref{mkm_full}) does not show bimodality in the metrics which most likely 
indicates that there is indeed no ``classical'' mode switching.

We note that we do not observe a monotonic variation in the profile across time unlike effects due to spin precession. The timescale of 
variation observed is also much lower compared to the typical timescales of profile evolution due to precession effects. \cite{Kramer_1999} argued that a propagation effect through the pulsar magnetosphere would most likely explain the changing components of the profile. This was reiterated by the study of \cite{Ramachandran_2003} where they analysed the polarisation properties of the pulsar. They observed a jump in the polarisation angle swing in a position coinciding with the leading component and suggested that the model proposed by \cite{Hibschman_2001} best fits the data. \cite{Liu_2015} suggested that the two components could arise from different emission heights but would require further investigation. While these models help explain the emission mechanism of this pulsar, observing and timing PSR~J1022+1001 with telescopes like MeerKAT and the upcoming SKA offers better sensitivity to study these subtle variations and further constrain the emission theories  previously proposed.

\section{Conclusions}

We were able to demonstrate results indicating that PSR~J1022+1001 has profile shape changes at long integration time scales. We modified methods suggested in preceding studies to make them more robust. We applied these methods not only on previously used data sets but also on new data from the latest Effelsberg pulsar backend. The EPOS data set specifically shows a higher degree of instability compared to the other data sets. We demonstrated that this is due to greater degree of noise in the back end chain of EPOS compared to PSRIX. The profile shape changes seen with CPSR2 show that instrumental errors cannot significantly change the pulse profile. We demonstrate that the calibration error variation is a tiny fraction of the total variations seen (up to maximum of 1 \emph{per cent}). Thus, there are additional instabilities that cannot be explained fully by miscalibration. With the CPSR2 and PSRIX data, we tested the effect of interstellar scintillation across frequency and showed that the scale of this variation can at most account for 25 \emph{per cent} of the observed variations.
The results strongly suggest that intrinsic emission factors are the major contribution to the observed profile variation.

\section*{Acknowledgements}

We would like to thank S. Johnston and W. van Straten for providing Jones matrix solution files for the CPSR2 data set. We thank Joris Verbiest, Nicolas Cabalero and Eleni Graikou for helping in the PSRIX data reduction. We would also like to thank Yanjun Guo for assisting to assess the quality of the data related to this work. We thank the referee for critical inputs and suggestions that provided weight to the text and the analysis methods. This work was partly based on observations with the 100-m telescope of the MPIfR (Max-Planck-Institut f\"ur Radioastronomie) at Effelsberg. The Parkes radio telescope is part of the Australia Telescope National Facility which is funded by the Australian Government for operation as a National Facility managed by CSIRO. The CSIRO Data Access Portal (DAP) is an institutional data repository supporting the publication of data and software for its organisation, CSIRO, Australia's national science agency. The aim of the CSIRO Data Access Portal is to provide reliable, long-term access to managed digital resources for CSIRO. It complies with the Force 11 Data Citation Principles.
An attribution statement and a DOI are assigned to all data deposited and made publicly available in the CSIRO Data Access Portal. DOIs are minted via the Australian National Data Service (ANDS) and DataCite. The CSIRO DAP complies with ANDS service policies. Data is stored persistently within the repository with full metadata including unique identifiers and a licence. It is mirrored by at least two separate data centres by default. Data and metadata are versioned.

\section*{Data Availability}

The data underlying this article will be shared on reasonable request to the corresponding author.

%%%%%%%%%%%%%%%%%%%%%%%%%%%%%%%%%%%%%%%%%%%%%%%%%%



\bibliographystyle{mnras}
\bibliography{1022} % if your bibtex file is called example.bib

%\begin{thebibliography}{99}
%\bibitem[\protect\citeauthoryear{Kramer et al.}{2006}]{kramer1}
%Science, Volume 314, Issue 5796, pp. 97-102 
%\end{thebibliography}

%%%%%%%%%%%%%%%%%%%%%%%%%%%%%%%%%%%%%%%%%%%%%%%%%%

%%%%%%%%%%%%%%%%% APPENDICES %%%%%%%%%%%%%%%%%%%%%

 \appendix

 \section{Original method analysis}
\label{appendix}

\begin{figure}
    \centering
    \includegraphics[scale=0.45]{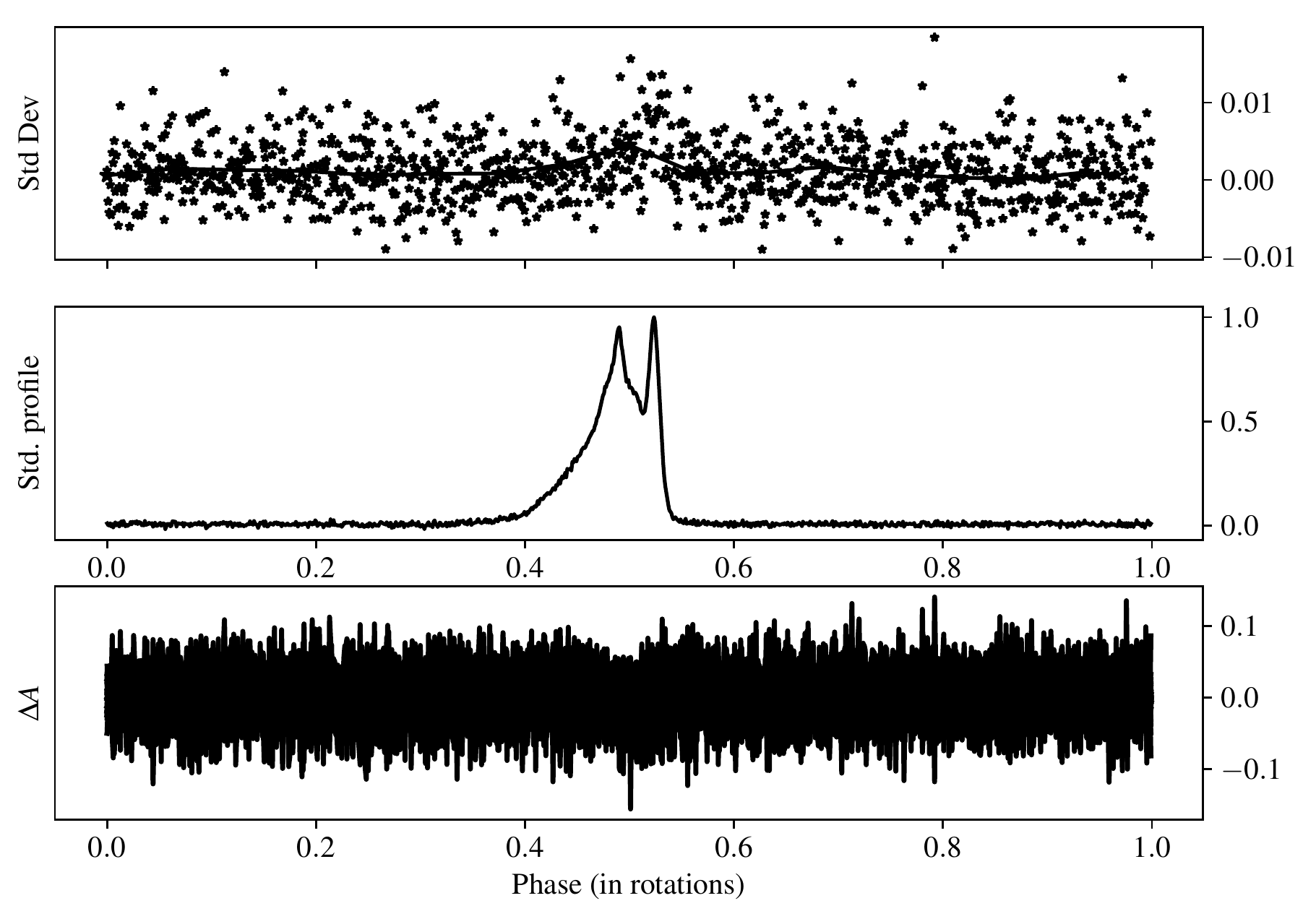}
    \caption{Results from applying the original  Hotan et al. method on CPSR2 (uncalibrated) data with 5 min long integrations as used by Hotan et al.
    }
    \label{cpsr2_ohm_5min}
\end{figure}

\begin{figure}
    \centering
    \includegraphics[scale=0.45]{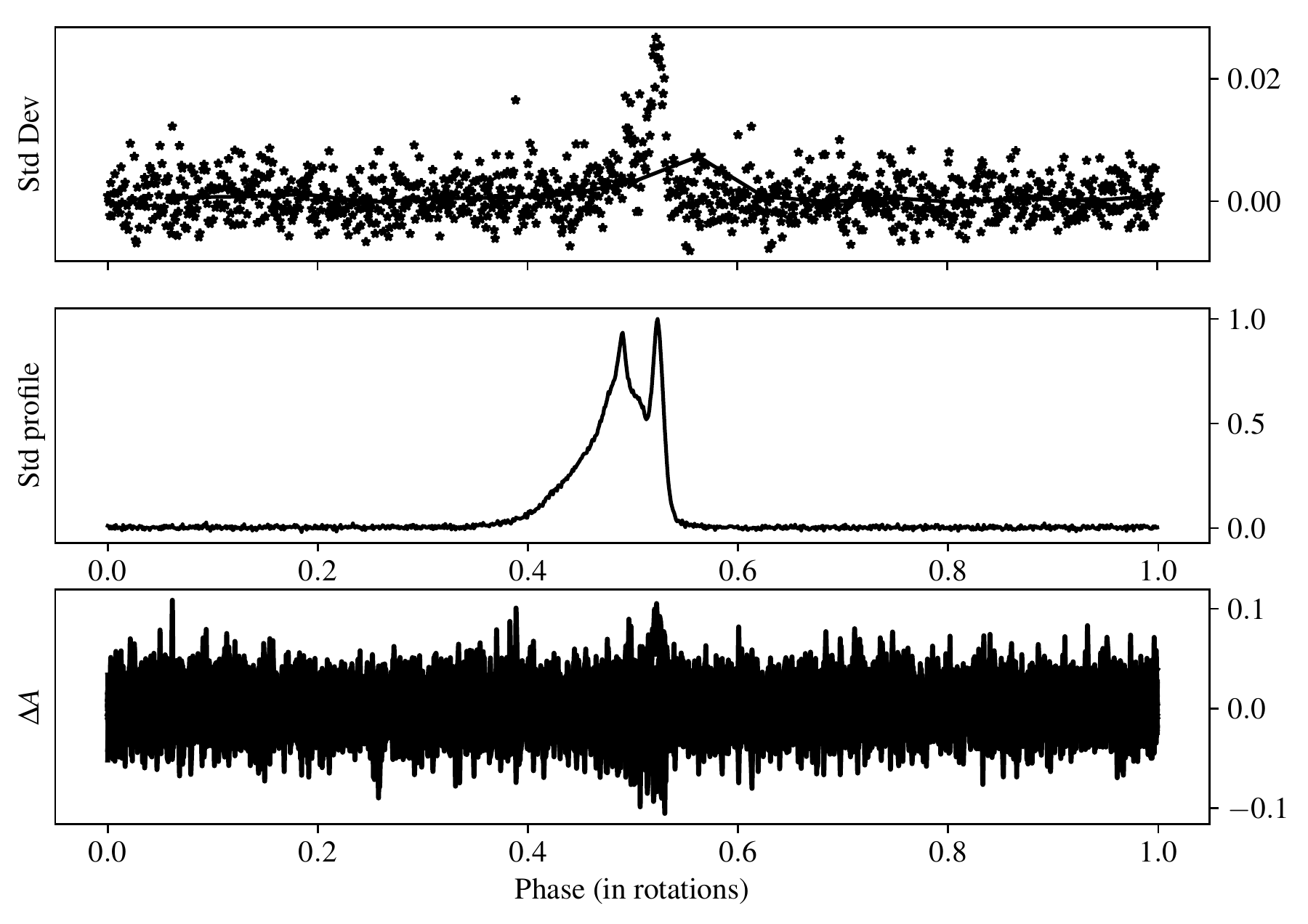}
    \caption{Results from applying the original  Hotan et al. method on CPSR2 (uncalibrated) data with 20-30 min long integrations.}
    \label{cpsr2_ohm_uncal}
\end{figure}

\begin{figure}
    \centering
    \includegraphics[scale=0.45]{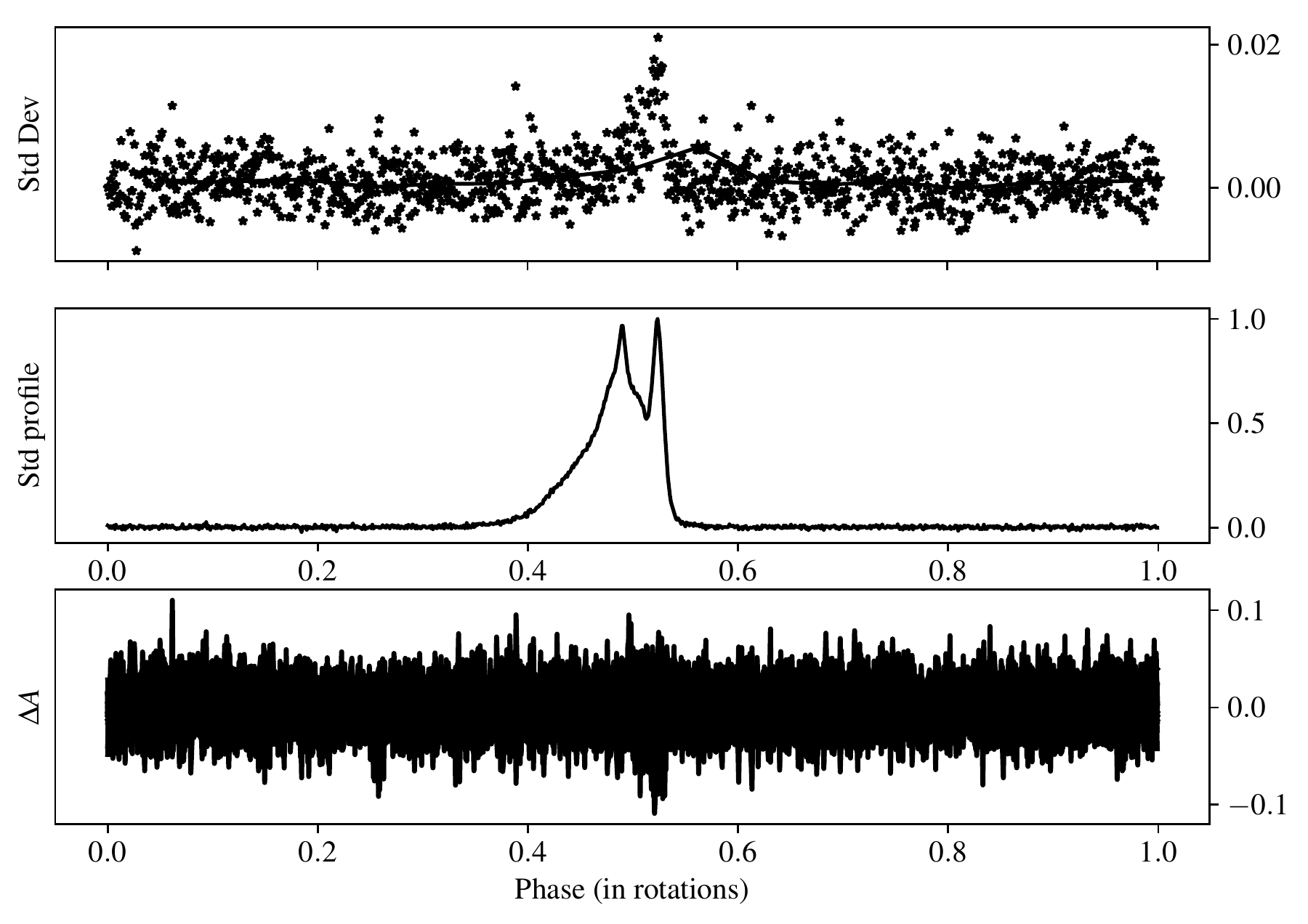}
    \caption{Results from applying the original  Hotan et al. method on CPSR2 (calibrated) data with 20-30 min long integrations.}
    \label{cpsr2_ohm_cal}
\end{figure}

We present here the results of applying the method described by \cite{2004MNRAS.355..941H} on CPSR2 data. Results described in figure \ref{cpsr2_ohm_5min} is obtained by applying the method on 5 min integration lengths of data as used by \cite{2004MNRAS.355..941H}. 
The bottom panel shows superposed 
difference profiles of the pulsar with a standard high S/N reference profile. The middle panel shows the mean profile. The top panel represents the standard deviation of per phase bin (marked as stars) in the difference profiles displayed in the bottom panel. The mean value of the standard deviation for every 64 bins is connected
by lines to show trends in the data.
We extended the analysis to the expanded CPSR2 dataset (calibrated and uncalibrated) with 20-30 min integration lengths. Figures \ref{cpsr2_ohm_uncal} and \ref{cpsr2_ohm_cal} show the results for uncalibrated and calibrated data respectively which show a much more significant trend for variability as compared to the 5 min data. 

% If you want to present additional material which would interrupt the flow of the main paper,
% it can be placed in an Appendix which appears after the list of references.

%%%%%%%%%%%%%%%%%%%%%%%%%%%%%%%%%%%%%%%%%%%%%%%%%%

% Don't change these lines
\bsp	% typesetting comment
\label{lastpage}
\end{document}